\documentclass[aps,prd,twocolumn,amsmath]{revtex4}

\usepackage{epsfig}
\usepackage{slashed}

\def\al{\alpha}
\def\be{\beta}
\def\ga{\gamma}
\def\de{\delta}
\def\ep{\epsilon}

\def\ze{\zeta}
\def\et{\eta}
\def\th{\theta}

\def\ka{\kappa}

\def\rh{\rho}

\def\si{\sigma}

\def\ph{\phi}

\def\ch{\chi}
\def\ps{\psi}
\def\om{\omega}
\def\Ga{\Gamma}

\def\cA{{\cal A}}

\def\cl{{\cal L}}

\def\cM{{\cal M}}

\def\mn{{\mu\nu}}

\def\fr#1#2{{{#1} \over {#2}}}
\def\half{{\textstyle{1\over 2}}}
\def\frac#1#2{{\textstyle{{#1}\over {#2}}}}

\def\vev#1{\langle {#1}\rangle}

\def\lsim{\mathrel{\rlap{\lower4pt\hbox{\hskip1pt$\sim$}}
    \raise1pt\hbox{$<$}}}
\def\gsim{\mathrel{\rlap{\lower4pt\hbox{\hskip1pt$\sim$}}
    \raise1pt\hbox{$>$}}}

\def\pt#1{\phantom{#1}}
\def\ol#1{\overline{#1}}

\def\etal{{\it et al.}}

\def\Re{\hbox{Re}\,}
\def\Im{\hbox{Im}\,}

\def\lrpartial{\raise 1pt\hbox{$\stackrel\leftrightarrow\partial$}}

\def\lrprtmu{\stackrel{\leftrightarrow}{\partial_\mu}}
\def\lrprtnu{\stackrel{\leftrightarrow}{\partial^\nu}}

\def\lrDnu{\stackrel{\leftrightarrow}{D^\nu}}

\newcommand{\beq}{\begin{equation}}
\newcommand{\eeq}{\end{equation}}
\newcommand{\bea}{\begin{eqnarray}}
\newcommand{\eea}{\end{eqnarray}}
\newcommand{\rf}[1]{(\ref{#1})}

\def\mi{m_i}
\def\mib{\ol m_i}
\def\gw{g}
\def\gs{g_{\rm s}}

\begin{document}

\title{Lorentz and CPT Violation
in Top-Quark Production and Decay}

\author{Micheal S.\ Berger, V.\ Alan Kosteleck\'y, and Zhi Liu}
\affiliation{Physics Department, Indiana University,
Bloomington, IN 47405, U.S.A.} 

\date{IUHET 588, September 2015}

\begin{abstract}

The prospects are explored for testing Lorentz and CPT symmetry
in the top-quark sector.
We present the relevant Lagrange density,
discuss physical observables,
and describe the signals to be sought in experiments.
For top-antitop pair production 
via quark or gluon fusion with subsequent semileptonic or hadronic decays,
we obtain the matrix element in the presence of Lorentz violation
using the narrow-width approximation. 
The issue of testing CPT symmetry in the top-quark sector
is also addressed.
We demonstrate that single-top production and decay
is well suited to a search for CPT violation,
and we present the matrix elements for 
single-top production in each of the four tree-level channels.
Our results are applicable to searches for Lorentz violation 
and studies of CPT symmetry in collider experiments,
including notably high-statistics top-antitop 
and single-top production at the Large Hadron Collider.

\end{abstract}

\maketitle

\section{Introduction}

The 1995 discovery of the top ($t$) quark at the Fermilab Tevatron 
\cite{d0cdf}
opened a new era for investigation
of the Standard Model (SM) of particle physics.
While experimental observations initially involved comparatively few events,
the advent of the Large Hadron Collider (LHC) has radically changed
the prospects for physics analyses involving the top quark.
Indeed,
the LHC can reasonably be viewed as a top-quark factory,
since it is expected to produce several million
single-top or single-antitop events
and even more top-antitop pairs
over the next few years,
with ultimately another order of magnitude 
produced during the lifetime of the machine
\cite{wb}.
The accompanying plethora of top-quark data,
which is unlikely to be matched at another collider 
in the foreseeable future,
offers a remarkable opportunity for precision measurements
using the heaviest elementary fermion in the SM.

Most of the precision measurements involving the top quark
that have been undertaken to date
either attempt to verify basic predictions of the SM
or search for new physics from models
constructed within a conventional field-theoretic context.
However,
the high statistical power provided by the LHC dataset
provides the opportunity to use top-quark physics
to investigate profound issues involving the validity of underlying features 
of quantum field theory and the SM in the third generation.
In the present work,
we explore the prospects for studying 
the foundational Lorentz and CPT symmetries of the SM
at the scale of the top quark.

Interest in precision tests of spacetime symmetries 
has grown significantly in recent years,
following the observation that tiny violations of Lorentz and CPT invariance
could arise naturally 
in an underlying unified theory such as strings
and be described at accessible energy scales 
using effective field theory
\cite{ksp}.
At this stage,
numerous experiments using methods from a variety of subfields
have sought evidence for Lorentz and CPT violation
\cite{tables},
but to date only one measurement investigating Lorentz symmetry
in the top-quark sector has been performed
\cite{D0top}. 
Some theoretical motivation for top-quark studies
comes from the notion that Lorentz violation
in a complete spacetime theory involving gravity 
is expected to be spontaneous rather than explicit,
as the latter is generically incompatible 
with conventional Riemann geometry or technically unnatural
\cite{ak,rb}. 
Supposing that Lorentz violation indeed arises spontaneously
through the vacuum expectation value of one or more tensor fields
in the underlying theory,
then the sizes of low-energy effects 
are governed in part by the couplings to these fields.
If the latter follow the familiar pattern of Yukawa couplings in the SM
determining the hierarchy of quark masses, 
then Lorentz- and CPT-violating effects might naturally be expected
to be largest for the top quark.
Moreover,
since the top quark decays before hadronization,
it offers a unique arena for studying Lorentz and CPT symmetry
in essentially free quarks.
In any case,
independent of deeper potential theoretical motivations,
as a matter of principle 
it is of interest to establish the laws of relativity
for the top quark on as firm a footing as possible.
 
The comprehensive realistic effective field theory
for Lorentz and CPT violation,
called the Standard-Model Extension (SME),
contains by construction the SM coupled to General Relativity
along with all possible operators for Lorentz violation
\cite{ak,ck}.
In realistic effective field theories
CPT violation is accompanied by Lorentz violation 
\cite{owg},
so the SME also describes general CPT violation.
A Lorentz-violating term in the Lagrange density of the SME
is an observer scalar density
formed by contracting a Lorentz-violating operator
with a coefficient for Lorentz violation 
that controls the size of the associated effects.
The operators can be classified systematically
using their mass dimension $d$,
with arbitrarily large values of $d$ appearing.
The restriction of the SME to include 
only Lorentz-violating operators with $d\leq 4$,
called the minimal SME,
is a renormalizable theory in Minkowski spacetime.
The SME provides a realistic and calculable framework
for analyses of experimental data
searching for deviations from Lorentz and CPT invariance 
\cite{reviews}.

During recent years,
many measurements have been performed
of fundamental properties of the top quark,
including its mass
\cite{topmass},
its charge
\cite{topcharge},
and its width
\cite{topwidth}. 
However,
to date the sole search for Lorentz violation in the top-quark sector 
was performed by the D0 Collaboration
\cite{D0top}
using data from the Fermilab Tevatron Collider 
and the theoretical formalism of the SME.
The production of $t$-$\ol t$ pairs 
at the Tevatron is dominated by quark fusion,
and the D0 Collaboration studied data
corresponding to an integrated luminosity of 5.3 fb$^{-1}$
for processes with the $t$-$\ol t$ pairs 
decaying into leptonic and jet final states.
These processes are primarily sensitive
to certain dimensionless SME coefficients for CPT-even Lorentz violation,
and the investigation constrains possible Lorentz violation 
involving these coefficients to below about the 10\% level.
The substantially greater statistical power available at the LHC
offers the opportunity to improve significantly on this study.
However,
at the LHC the primary production mechanism 
for $t$-$\ol t$ pairs is gluon fusion,
for which the matrix elements are different
and more involved than those for quark fusion. 
One goal of the present work is to present the essential theory
appropriate for $t$-$\ol t$ production by gluon fusion.

Another interesting issue is the extent to which CPT symmetry 
is respected by the top quark.
Since CPT violation comes with Lorentz violation
in realistic effective field theory
\cite{ck,owg},
studies of CPT violation necessarily involve observables 
that change with energy and orientation.
No experimental investigations of CPT symmetry for the top quark
in this context have been performed to date.
In this work,
we partially address this gap in the literature
by demonstrating that studies of single-top or single-antitop production 
at the LHC provide the basis for a search for CPT violation. 

The structure of this paper is as follows.
We begin in Sec.\ \ref{Theory}
by establishing the basic theory used in this work.
The relevant parts of the SME Lagrange density are provided,
the physical observables are identified,
and the types of signals of relevance are discussed.
We then turn in Sec.\ \ref{Top-antitop pair production} 
to top-antitop pair production,
where we present the matrix element for production and decay.
The Lorentz-invariant result is given,
followed by a demonstration that pair production
is a CPT-even process.
We give the explicit amplitudes 
for Lorentz-violating pair production and decay 
both via quark fusion,
which was the dominant process for the D0 analysis,
and via gluon fusion,
which dominates at the LHC.
In Sec.\ \ref{Single-top production},
we address CPT violation in the top-quark sector,
showing that single-top production offers access to CPT observables.
Four tree-level channels play a role,
and we derive the matrix elements for each. 
Details of the spin sum required
for calculations of the single-top matrix elements
are relegated to appendix \ref{appendix}.
We conclude with a summary and discussion in Sec.\ \ref{Summary}.
Throughout this work,
our conventions are those adopted in Ref.\ \cite{ck}.

\section{Theory}
\label{Theory}

This section provides some theoretical comments of relevance
to the derivations in the remainder of the paper.
We present the portion of the SME Lagrange density 
applicable to the top-quark searches studied here,
discuss the issues of field redefinitions and physical observables,
and offer some observations about generic signals 
that could be sought in experimental analyses.

\subsection{SME Lagrange density for the top quark}
\label{SME Lagrange density for the top}

In this paper,
our focus is on the top-quark sector in the minimal SME.
The part of the SME Lagrange density
involving Lorentz and CPT violation in the top-quark sector 
can be extracted from Ref.\ \cite{ck}.
Denoting the left-handed quark doublets by $Q_A$ 
and the right-handed charge-2/3 singlets as $U_A$, 
the relevant piece of these equations 
describing CPT-even Lorentz violation is
\bea
\cl^{\rm CPT+} &\supset& 
\half i (c_Q)_{\mu\nu AB} \ol{Q}_A \ga^{\mu} \lrDnu Q_B
\nonumber\\ &&
+ \half i (c_U)_{\mu\nu AB} \ol{U}_A \ga^{\mu} \lrDnu U_B 
\nonumber\\
&&
- \half (H_U)_{\mu\nu AB} \ol{Q}_A \ph^c \si^{\mu\nu} U_B 
+ {\rm h.c.},
\label{lorviolq}
\eea
where $D_\mu$ is the gauge-covariant derivative
and $\ph$ is the Higgs field.
The piece governing CPT-odd Lorentz violation is
\beq
\cl^{\rm CPT-} \supset 
- (a_Q)_{\mu AB} \ol{Q}_A \ga^{\mu} Q_B
- (a_U)_{\mu AB} \ol{U}_A \ga^{\mu} U_B .
\label{cptviolq}
\eeq
The various coefficients in these equations
determine the size of the Lorentz violation.
The dimensionless coefficients $c_{\mu\nu AB}$ 
are traceless in spacetime indices $\mu, \nu$
and are hermitian in generation indices $A,B$,
while the dimensionless coefficients $H_{\mu\nu AB}$ 
are antisymmetric in spacetime indices $\mu, \nu$.
The coefficients $a_{\mu AB}$ have dimensions of mass
and are hermitian in generation indices $A,B$.

In this work,
which focuses on the top quark,
we assume for definiteness and simplicity 
that the only relevant Lorentz and CPT violation
involves the third generation,
so that $A=B=3$.
A more general treatment would also be of interest
but lies outside our present scope.
The coefficients of relevance here are therefore 
$(c_Q)_{\mu\nu 33}$, 
$(c_U)_{\mu\nu 33}$, 
$(H_U)_{\mu\nu 33}$,
$(a_Q)_{\mu 33}$, 
and $(a_U)_{\mu 33}$. 
The first three control CPT-even operators,
while the last two control CPT-odd ones.
All coefficients affect the propagator for the top-quark field $t$,
while $(c_Q)_{\mu\nu 33}$ and $(a_Q)_{\mu 33}$ also affect
the propagator for the bottom-quark field $b$,
and $(c_Q)_{\mu\nu 33}$ affects the $t$-$b$-$W$ vertex as well.
For convenience in what follows,
we introduce the abbreviated notation
\bea
(a_L)_{\mu} &=& (a_Q)_{\mu 33}, 
\quad
(a_R)_{\mu} = (a_U)_{\mu 33}, 
\nonumber\\
(c_L)_{\mu\nu} &=& (c_Q)_{\mu\nu 33}, 
\quad
(c_R)_{\mu\nu} = (c_U)_{\mu\nu 33}, 
\nonumber\\
H^\prime_{\mu\nu} &=& \vev{\ph} (H_U)_{\mu\nu 33},
\quad
\widetilde H^\prime{}^{\mu\nu} = 
\half \ep^{\mu\nu\rh\si} H^\prime_{\rh\si},
\eea
where $\vev{\ph}$ is the Higgs expectation value.
It is also useful to define certain coefficient combinations as
\bea
a_\mu &=& \half[ (a_L)_{\mu} + (a_R)_{\mu} ],
\quad
b_\mu = \half [ (a_L)_{\mu} - (a_R)_{\mu} ],
\nonumber\\
c_{\mu\nu} &=& \half [ (c_L)_{\mu\nu} + (c_R)_{\mu\nu} ],
\quad
d_{\mu\nu} = \half [ (c_L)_{\mu\nu} - (c_R)_{\mu\nu} ],
\nonumber\\
H_{\mu\nu} &=& 
\Re{H^\prime_{\mu\nu}}
- \Im{\widetilde H^\prime_{\mu\nu}}.
\label{LRdefs}
\eea
In these expressions,
all coefficients are real except for 
$(H_U)_{\mu\nu 33}$, $H^\prime_{\mu\nu}$,
and its dual $\widetilde H^\prime{}^{\mu\nu}$,
which may be complex.

For our purposes,
the relevant part $\cl_{t,b}^{\rm SM}$
of the matter Lagrange density for the conventional SM
involves the $t$ and $b$ quark fields,
their electroweak interactions with the $W^\pm_\mu$ bosons,
and their strong interactions with the SU(3)-adjoint matrix $G_\mu$ of gluons.
In what follows,
the left- and right-handed fermion fields are defined by 
\beq
\ps_L \equiv \half (1-\ga_5)\ps,
\quad
\ps_R \equiv \half (1+\ga_5)\ps,
\eeq
as usual.
Using this notation,
\bea
\cl_{t,b}^{\rm SM} &=& 
\half i \ol{t} \ga^\mu \lrprtmu t 
- m_t \ol t t
+ \half i \ol{b} \ga^\mu \lrprtmu b 
- m_b \ol b b
\nonumber\\ 
&&
+ (\fr {\gw V_{tb}} {\sqrt{2}} W_\mu^- \ol b_L \ga^\mu t_L + {\rm h.c.})
\nonumber\\ 
&&
+ \gs (\ol t \ga^\mu G_\mu t  + \ol b \ga^\mu G_\mu b ) ,
\label{smquarks}
\eea
where $m_t$ and $m_b$ are the masses of the top and bottom quarks,
respectively,
$\gw$ is the electroweak coupling constant,
$V_{tb}$ is an element of the Cabibbo-Kobayashi-Maskawa (CKM) matrix,
and $\gs$ is the strong coupling constant.
The SME corrections involving CPT-even Lorentz violation
can be extracted from Eq.\ \rf{lorviolq}
and written in various equivalent forms,
\bea
\cl^{\rm CPT+}_{t,b} &=& 
\half i (c_L)_{\mu\nu} \ol t_L \ga^\mu \lrprtnu t_L 
+ \half i (c_R)_{\mu\nu} \ol t_R \ga^\mu \lrprtnu t_R 
\nonumber \\ &&
+ \half i (c_L)_{\mu\nu} \ol b_L \ga^\mu \lrprtnu b_L
\nonumber \\ &&
+ (\fr {\gw V_{tb}} {\sqrt{2}} (c_L)_{\mu\nu} W^{-\nu} 
\ol b_L \ga^\mu t_L + {\rm h.c.})
\nonumber \\ &&
- \half H_{\mu\nu} \ol t_L \si^{\mu\nu} t_R 
- \half H_{\mu\nu} \ol t_R \si^{\mu\nu} t_L 
\nonumber \\ &&
+ \gs c_\mn (\ol t \ga^\mu G^\nu t  + \ol b \ga^\mu G^\nu b ) ,
\nonumber\\
&=&
\half i c_{\mu\nu} \ol t \ga^\mu \lrprtnu t 
+ \half i d_{\mu\nu} \ol t \ga^5 \ga^\mu \lrprtnu t 
\nonumber \\ &&
+ \half i (c_L)_{\mu\nu} \ol b_L \ga^\mu \lrprtnu b_L
\nonumber \\ &&
+ (\fr {\gw V_{tb}} {\sqrt{2}} (c_L)_{\mu\nu} W^{-\nu} 
\ol b_L \ga^\mu t_L + {\rm h.c.})
\nonumber \\ &&
- \half H_{\mu\nu} \ol t \si^{\mu\nu} t 
+ \gs c_\mn (\ol t \ga^\mu G^\nu t  + \ol b \ga^\mu G^\nu b ) .
\qquad
\label{lorviol}
\eea
Similarly,
the CPT-odd terms 
obtained from Eq.\ \rf{cptviolq}
can be written 
\bea
\cl^{\rm CPT-}_{t,b} 
&=& 
- (a_L)_\mu \ol t_L \ga^\mu t_L 
- (a_R)_\mu \ol t_R \ga^\mu t_R 
\nonumber\\ &&
- (a_L)_\mu \ol b_L \ga^\mu b_L 
\nonumber\\
&=&
- a_\mu \ol t \ga^\mu t 
- b_\mu \ol t \ga^5 \ga^\mu t 
- (a_L)_\mu \ol b_L \ga^\mu b_L 
\nonumber\\
&=&
- a_\mu \ol t \ga^\mu t 
- b_\mu \ol t \ga^5 \ga^\mu t 
\nonumber \\ &&
- \half (a_L)_\mu \ol b \ga^\mu b 
- \half (a_L)_\mu \ol b \ga^5 \ga^\mu b .
\label{cptviol}
\eea
In subsequent sections,
the above expressions are used to derive
the matrix elements for top-antitop production and decay
and to explore the prospects for studying CPT violation 
in single-top production.

\subsection{Observables}
\label{Observables}

For top-quark production and decay,
the Lorentz-violating terms listed above
can affect Feynman diagrams through the production vertices,
the $t$ and $\ol t$ propagators,
the decay vertices,
and the $b$ and $\ol b$ propagators. 
At leading order,
each contribution from Lorentz violation
arises as an insertion on a propagator or a vertex.
The matrix element for a Lorentz-violating process
can then be computed from the Feynman diagrams
in the usual way,
except perhaps for some technical issues involving external legs
\cite{extleg}.
However,
only a subset of the Lorentz-violating insertions 
lead to physically observable effects.
Some terms can be absorbed into unobservable phases in the fields,
choices for the spacetime coordinates, 
or redefinitions of the spinor basis
\cite{ck,ak,akmm,redefref}.
These terms therefore can be expected to cancel
in matrix elements.
To minimize calculations,
it is useful to identify relevant terms beforehand. 
In this subsection,
we outline the procedure for this.

For the analysis,
one could in principle work with the SME 
prior to the SU(2)$\times$U(1) breaking,
at the level of Eqs.\ \rf{lorviolq} and \rf{cptviolq}.
The relevant spinor fields subject to redefinitions
would then be $Q_A$ and $U_A$,
but care must be taken because 
the two components of each quark doublet can play independent roles.
Here,
we work instead with the terms \rf{lorviol} and \rf{cptviol}
belonging to the SME Lagrange density 
after the SU(2)$\times$U(1) breaking,
for which the relevant spinor fields are 
$t_L$, $t_R$, $b_L$, $b_R$.
It suffices for present purposes to consider
the observability of coefficients at leading order.

Consider first the terms \rf{lorviol} for CPT-even Lorentz violation.
Each coefficient of the $c_{\mu\nu}$ and $d_{\mu\nu}$ type
is a sum of three observer Lorentz irreducible pieces:
a trace, a symmetric part, and an antisymmetric part.
The traces of these coefficients
are irrelevant for our purposes
because they are Lorentz invariant
and can be absorbed into overall normalizations 
of the fields,
so they can be set to zero without loss of generality.
Also,
the symmetric parts of these coefficients are all physically observable
in principle.
Specifying the particle sector used to define the Minkowski metric
normally removes one symmetric coefficient of the $c_{\mu\nu}$ type,
but in the present instance we have already tacitly made such a choice 
by assuming that the light quark sectors are conventional.
In contrast,
only some of the antisymmetric parts of the coefficients 
of the $c_{\mu\nu}$ and $d_{\mu\nu}$ type 
can be physically observable
due to the possibility of field redefinitions
amounting to a choice of basis in spinor space.

As an explicit example,
consider the redefinition 
$t_L\to (1 + i v_{\mu\nu} \si^{\mu\nu})t_L$
where $v_{\mu\nu}$ is constant,
and perform the same redefinition on $t_R$, $b_L$, $b_R$.
We remark that although this redefinition
superficially resembles an infinitesimal Lorentz transformation
under which the Lagrange density is invariant,
here only a subset of the fields are involved 
and so the form of the Lagrange density changes.
Under this redefinition,
the kinetic term for $t$ in Eq.\ \rf{smquarks}
generates a term of the $c_{\mu\nu}$ type for $t$,
while the $t$ mass term is invariant.
Similarly,
the $b$ kinetic term
generates a term of the $c_{\mu\nu}$ type for $b$,
and the $b$ mass term is invariant.
The $W^-_\mu$ interaction term
produces an interaction term involving
$(c_L)_{\mu\nu}$,
while the gluon couplings are invariant.
These results imply that all the antisymmetric parts
of $c_{\mu\nu}$ appearing in Eq.\ \rf{lorviol}
can be absorbed by a suitable choice of $v_{\mu\nu}$,
at the cost of introducing a term of the form
$i (c_L)_{\mu\nu} \ol b_R \ga^\mu \lrprtnu b_R$
for the $b_R$ field.
It then follows,
for example,
that the antisymmetric part of $c_{\mu\nu}$ 
is irrelevant for top-quark production,
and also that any effects on top decays
can be attributed to the propagator terms
for $b$ and $\ol b$.
A similar line of reasoning reveals that
the antisymmetric part of $d_{\mu\nu}$
can be absorbed into $H_{\mu\nu}$ and elsewhere in the Lagrange density,
so it can therefore be viewed as irrelevant for present purposes as well.

Next,
consider the terms \rf{cptviol} for CPT-odd Lorentz violation.
Suppose first that $t_L$ is redefined 
by an unobservable position-dependent phase,
$t_L\to \exp(- i v_\mu x^\mu)t_L$,
where $v_\mu$ is constant.
The mass term for $t$ in Eq.\ \rf{smquarks} 
remains unaffected provided $t_R$ is redefined the same way.
The kinetic term for $t$ 
in the SM Lagrange density 
then generates a term of the $a_\mu$ type for $t$.
The $W_\mu^-$-interaction term is invariant if $b_L$
is redefined in this way,
and the mass term for $b$ is unchanged if $b_R$
is too.
The kinetic term for $b$ 
in the SM Lagrange density 
then generates a term of the $a_\mu$ type for $b$.
These results imply that the freedom in choosing $v_\mu$ allows,
for example,
removing the term $-\half (a_L)_\mu \ol b \ga^\mu b$ 
from the last line of Eq.\ \rf{cptviol}
without loss of generality.
A useful option is to choose 
$v_\mu$ to cancel $(a_L)_\mu$
in the first and third terms of the first line of Eq.\ \rf{cptviol},
leaving the physically equivalent terms 
\bea
\cl^{\rm CPT-}_{t,b} &\equiv& 
[(a_L)_\mu - (a_R)_\mu] \ol t_R \ga^\mu t_R 
+ (a_L)_\mu \ol b_R \ga^\mu b_R 
\nonumber\\
&=&
b_\mu \ol t \ga^\mu t
- b_\mu \ol t \ga^5 \ga^\mu t
\nonumber\\
&&
+ \half (a_L)_\mu \ol b \ga^\mu b
- \half (a_L)_\mu \ol b \ga^5 \ga^\mu b
\label{cptviolqequiv}
\eea
in which only right-handed fields appear.

With the above choices,
the description of CPT violation in the top-quark sector
is reduced to considerations of insertions
involving only right-handed fields,
thereby simplifying both practical calculations
and physical intuition.
For the latter,
for example,
with the effects of CPT violation limited to the $t$ propagator
according to Eq.\ \rf{cptviolqequiv},
a top quark follows a geodesic in a pseudo-Finsler spacetime
\cite{finsler}. 
In a related vein,
we remark in passing that
no mass differences between $t$ and $\ol t$ appear,
a result in accordance with Greenberg's theorem
\cite{owg}
and also with expectations for CPT violation 
in realistic effective field theory,
where the antitop particle associated with the field $\ol t$ 
is defined as the antiparticle of $t$
and therefore by construction always has the same Lagrange-density mass.

Further simplifications affecting the observability of CPT violation
appear if one or more of the quark masses
can be neglected in a given process. 
For example,
when the $b$ mass $m_b$ is negligible
compared to the $b$ kinetic term,
then the field $b_R$ can be independently redefined
using a different phase,
$b_R\to \exp(- i v'_\mu x^\mu)b_R$,
while leaving unaffected the form of the SM Lagrange density
except for generating a term of the $a_\mu$ type for $b$. 
A suitable choice of $v'$ therefore can eliminate 
all $b$-quark terms in Eq.\ \rf{cptviolqequiv}
when $m_b$ is negligible.
Any observable CPT-violating effects on $t$ processes
must then arise from insertions on $t$-quark propagators.
Moreover,
if the $t$ mass $m_t$ itself is also negligible 
in a given experimental process,
then all $t$-quark terms in Eq.\ \rf{cptviolqequiv}
can be removed via another independent field redefinition
without changing the physics.
For this special limiting case 
and under the assumptions leading to 
Eqs.\ \rf{smquarks} and \rf{cptviol},
no top-quark CPT violation is observable at leading order.

\subsection{Signals}
\label{Signals}

Top-quark physics offers a rich variety of options 
for seeking Lorentz-invariant physics beyond the SM
\cite{beyondSM}.
However,
signals of Lorentz violation have unique features
that cannot be associated with Lorentz-invariant effects.
For example,
in a given inertial frame,
the presence of Lorentz violation
means that the properties of each quark 
depend on its direction of travel and its boost.
Moreover,
if the Lorentz violation includes CPT violation,
then the properties of the top and antitop can differ as well.
These features lead to distinctive experimental signals
that provide a basis for searches for Lorentz violation
in the top-quark sector,
as outlined in this subsection. 

For ease of comparison between experiments,
it is useful to introduce a standard inertial frame 
to report measurements of coefficients for Lorentz violation.
The canonical frame adopted in the literature 
is a Sun-centered frame
\cite{tables,akmm,sunframe}.
Cartesian coordinates $(T,X,Y,Z)$ in this frame 
are defined so that the $Z$ axis points along the
direction of the Earth's rotation,
while the $X$ axis points from the Earth to the Sun
at the vernal equinox 2000.
Unlike any Earth-based reference frame,
the Sun-centered frame can be taken as approximately inertial 
over a period of years.
Its alignment also means that the transformation between
the Sun-centered frame and a laboratory frame
is comparatively simple.
Suppose,
for example,
that cartesian coordinates $(t,x,y,z)$ in the laboratory 
are chosen such that the $x$-axis points south, 
the $y$-axis points east,
and the $z$-axis points vertically upwards.
Since the Earth rotates with sidereal frequency 
$\om_\oplus\simeq 2\pi$/(23 h 56 m)
the relationship mapping coefficients in the laboratory frame 
to those in the Sun-centered frame 
involves a time-dependent rotation $R^{jJ}$
between the two coordinate systems,
which is given explicitly as
\cite{akmm} 
\beq
R^{jJ}=\left(
\begin{array}{ccc}
\cos\ch\cos\om_\oplus T
&
\cos\ch\sin\om_\oplus T
&
-\sin\ch
\\
-\sin\om_\oplus T
&
\cos\om_\oplus T
&
0
\\
\sin\ch\cos\om_\oplus T
&
\sin\ch\sin\om_\oplus T
&
\cos\ch
\end{array}
\right),
\label{rotmat}
\eeq
where $\ch$ is the colatitude of the experiment.

For convenience of use,
the expressions derived in the sections below 
for the various matrix elements for top-quark production and decay
are expressed as observer Lorentz scalars,
so they can be evaluated in any observer frame.
For example,
the laboratory frame can be chosen as the specified observer frame,
and then the expressions for the matrix elements
can be evaluated with all coefficients and 4-momenta taken in that frame.
Since the coefficients in the laboratory frame
can be obtained from those in the Sun-centered frame
via the rotation \rf{rotmat},
experimental results can readily be reported directly
in terms of coefficients in the Sun-centered frame.
The reader is cautioned here to distinguish the observer Lorentz invariance,
which is merely an expression of coordinate independence of the physics,
from the physical particle Lorentz violation
arising through nonzero coefficients
\cite{ck}. 
For example,
in the presence of Lorentz violation,
physically boosting a particle of mass $m$ in any fixed observer frame
produces behavior governed by a modified dispersion relation
involving coefficients for Lorentz violation
\cite{kl01},
instead of the standard dispersion relation $p^2 = m^2$.

It is physically reasonable to take
the coefficients for Lorentz violation
as constant in the Sun-centered frame
\cite{aksidereal}.
Since the transformation to the laboratory frame
involves the time-dependent rotation \rf{rotmat},
the laboratory-frame coefficients vary with sidereal time.
Lorentz violation therefore can be expected 
to produce sidereal oscillations in the data,
with amplitudes and phases governed by the coefficients for Lorentz violation.
Most coefficients produce signals at the sidereal frequency $\om_\oplus$,
but the symmetric components of the coefficient $c_{\mu\nu}$ 
generate ones at the harmonic $2\om_\oplus$ as well.
Note also that the revolution of the Earth about the Sun 
introduces an extra time dependence in the laboratory-frame coefficients
with an annual periodicity.
However,
this time dependence is suppressed relative to the previous one 
by the Earth's boost $\be_\oplus \simeq 10^{-4}$,
which in practice reduces the experimental sensitivity
to annual variations 
to below the level of interest here.

The above considerations reveal that the data
for top-quark production and decay can be expected to contain
information in the amplitudes and phases 
of the the sidereal and twice-sidereal harmonics.
An analysis including many coefficients can be cumbersome,
but in practice considerable insight can be gained 
by allowing only one component of a coefficient for Lorentz violation 
to be nonzero at a time,
and extracting the sensitivity to it.
This simplified procedure is common practice in the field
\cite{tables}.
While it disregards possible interference
or cancellation of effects between coefficient components,
it does give a notion of the maximal sensitivity
achieved to each component. 
Evidently,
if a nonzero result is found for any component,
the question of possible interference would need to be revisited.
This type of analysis has been performed recently
in the context of top-antitop production by the D0 Collaboration
\cite{D0top},
who report limits of about 10\% on individual components
of the $c_{\mu\nu}$-type coefficients 
in the canonical Sun-centered frame.

In addition to studying sidereal signals,
which intrinsically include both CPT-even and CPT-odd Lorentz violation,
an analysis can also seek to isolate CPT-odd effects.
One method to achieve this is to work with a suitable asymmetry.
If the rate for a process is found to be $R$ 
and the rate for the CPT-conjugate process is $\ol R$,
then the asymmetry
\beq
\cA_{\rm CPT} \equiv \fr{R - \ol R}{R + \ol R}
\label{asym}
\eeq
provides a measure of CPT violation for that process.
Asymmetries of this type have been widely used
in the context of SME studies of CPT violation
in neutral-meson oscillations
\cite{aksidereal,akrvk},
where coefficients for various combinations of non-top quarks
have been experimentally constrained
using $K^0$, $D^0$, $B^0_d$, and $B^0_s$ mesons
\cite{KTEV, KLOE,FOCUS,BABAR,D0}.
The asymmetry $\cA_{\rm CPT}$ is proportional to coefficients
for CPT-odd Lorentz violation,
and typically it has both a constant term
and a term oscillating with sidereal time.
The experimental analysis therefore has several paths
available to extract different information.
Constructing the time-averaged asymmetry $\vev{\cA_{\rm CPT}}$,
for which oscillations average to zero over many sidereal days, 
permits constraints on coefficients enterng the constant term,
while studying the amplitudes and phases of the oscillations
provides independent measures of CPT symmetry.

\section{Top-antitop pair production} 
\label{Top-antitop pair production} 

In this section we discuss the matrix elements 
for top-antitop pair production in the presence of Lorentz violation.
We begin by outlining the Lorentz-invariant case,
including top-antitop pair production
both from quark fusion and from gluon fusion.
We then consider the Lorentz-violating situation,
showing that the process is CPT even
and deriving the matrix elements for both quark fusion and gluon fusion
at leading order in Lorentz violation.

\subsection{Lorentz-invariant case} 
\label{Lorentz-invariant case} 

\begin{figure}
\vskip -40pt
\begin{center}
\centerline{\psfig{figure=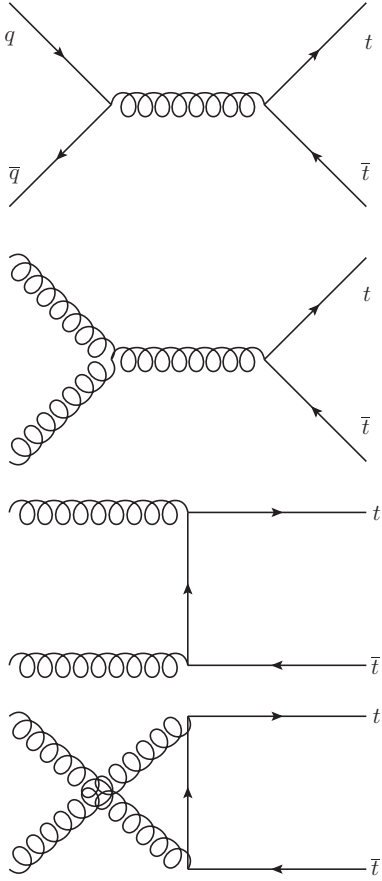,width=1.5\hsize}}
\vskip -100pt
\caption{
\label{fig1}
Tree-level Feynman diagrams for
$t$-$\ol t$ production via 
quark fusion ($q \ol q \to t \ol t$)
and gluon fusion ($gg \to t \ol t$). }
\end{center}
\end{figure}

Consider first the matrix element $\cM$ 
for the case without Lorentz violation.
The process of interest involves
the production of a $t$-$\ol t$ pair,
each component of which then decays.
In quark fusion the production is via a single gluon in the $s$ channel,
while in gluon fusion 
all three $s$, $t$, and $u$ channels contribute.
The tree-level diagrams of relevance are shown in Fig.\ \ref{fig1}.
Note that production via gluon fusion contributes 
only at the 15\% level at Tevatron energies 
\cite{knk},
while for production at LHC energies 
the situation is reversed
with gluon fusion dominating the process.
 
In the standard narrow-width approximation
\cite{nwa,Berdine:2007uv,Uhlemann:2008pm},
the squared modulus of $\cM$ can be written as the product of three parts,
\begin{eqnarray}
|\cM|^2=PF\ol{F}.
\label{m2}
\end{eqnarray}
The quantities $P$, $F$, and $\ol F$ 
represent the factors from the $t$-$\ol t$ pair production, 
the $t$ decay, 
and the $\ol t$ decay,
respectively.
Next,
we consider each factor in turn. 

The production factor $P$ is different for quark fusion and gluon fusion.
For quark fusion the production factor $P_{q\ol{q}}$
is given by 
\begin{eqnarray}
P_{q\ol{q}}&=&{\gs^4\over 9}(2-\beta^2\sin ^2\theta)
\nonumber\\
&=&{\gs^4\over {18E^4}}\Big [
(p_q\cdot p_t)(p_{\ol{q}}\cdot p_{\ol{t}})
+(p_q\cdot p_{\ol{t}})(p_{\ol{q}}\cdot p_t)
\nonumber\\
&&
\quad \qquad
+(p_q\cdot p_{\ol{q}})m_t^2\Big ],
\end{eqnarray}
where $\be$ is the common speed of the $t$ and $\ol{t}$ quarks 
in the production center-of-mass frame,
and $\th$ is the scattering angle.
The second equation above provides the expression
in terms of the four-momenta of the various particles involved,
with $E$ being the common energy of the quark or antiquark 
in the production center-of-mass frame, 
so that the usual Mandelstam variable
for the subprocess is $s=4E^2$.

For gluon fusion, 
the production factor $P_{2g}$
with color and polarization averaged and spins summed 
can be expressed as a sum of six contributions
arising from the three Feynman diagrams in the $s$, $t$, and $u$ channels
\cite{Gluck:1977,Combridge:1978kx,Ellis:1991qj},
\bea
P_{2g} &=& 
\ol{\sum}
\big(
|\cM_{ss}|^2
+|\cM_{tt}|^2
+|\cM_{uu}|^2
\nonumber\\
&&
\qquad
+|\cM_{st}|^2
+|\cM_{su}|^2
+|\cM_{tu}|^2
\big),
\eea
where
\bea
&&\ol{\sum}|\cM_{ss}|^2=
{{3\gs^4}\over 4}{{(t-m_t^2)(u-m_t^2)}\over {s^2}},
\nonumber \\
&&\ol{\sum}|\cM_{tt}|^2=
{{\gs^4}\over 6}{{(t-m_t^2)(u-m_t^2)-2m_t^2(t+m_t^2)}\over {(t-m_t^2)^2}},
\nonumber \\
&&\ol{\sum}|\cM_{uu}|^2=
{{\gs^4}\over 6}{{(u-m_t^2)(t-m_t^2)-2m_t^2(u+m_t^2)}\over {(u-m_t^2)^2}},
\nonumber \\
&&\ol{\sum}|\cM_{st}|^2=
{{3\gs^4}\over 8}{{(t-m_t^2)(u-m_t^2)+m_t^2(u-t)}\over {s(t-m_t^2)}},
\nonumber \\
&&\ol{\sum}|\cM_{su}|^2=
{{3\gs^4}\over 8}{{(u-m_t^2)(t-m_t^2)+m_t^2(t-u)}\over {s(u-m_t^2)}},
\nonumber \\
&&\ol{\sum}|\cM_{tu}|^2=
-{{\gs^4}\over {24}}{{m_t^2(s-4m_t^2)}\over {(t-m_t^2)(u-m_t^2)}}.
\eea
In these expressions,
$s$, $t$, and $u$ are the usual Mandelstam variables,
\bea
s&=&(p_1+p_2)^2=(p_t+p_{\ol t})^2
\nonumber \\
&=& 2 p_1 \cdot p_2 = 2 p_t \cdot p_{\ol t} + 2 m_t^2,
\nonumber \\
t&=&(p_2-p_t)^2=(p_1-p_{\ol t})^2
\nonumber \\
&=& -2 p_2 \cdot p_t + m_t^2 = -2 p_1 \cdot p_{\ol t} + m_t^2,
\nonumber \\
u&=&(p_1-p_t)^2=(p_2-p_{\ol t})^2
\nonumber \\
&=& -2 p_1 \cdot p_t + m_t^2 = -2 p_2 \cdot p_{\ol t} + m_t^2,
\label{mandelstam}
\eea
where $p_1$, $p_2$ are the 4-momenta of the two gluons.
Each of the six expressions above represents 
either the squared modulus of an individual diagram
or the interference between different channels.
The calculation uses the trick of modifying the $s$-channel diagram 
to remove the contribution of the unphysical gluon polarizations
\cite{Combridge:1978kx,Feynman:1963ax,Georgi:1978kx}. 
If instead the unphysical polarizations are handled 
by including the contribution from Fadeev-Popov ghosts 
in the squared matrix elements, 
then the individual expressions above differ 
but their sum remains unchanged. 

Within the narrow-width approximation,
the decay factors $F$, $\ol F$
are independent of the production mechanism.
Suppose for definiteness that the $t$ decays leptonically as
\beq
t\to W^+b\to \ol{l}\nu b
\label{tdecay}
\eeq
while the $\ol t$ decays hadronically as
\beq
\ol{t}\to W^-\ol{b}\to q\ol{q}' \ol{b}.
\label{tbardecay}
\eeq
The factor $F$ in Eq.\ \rf{m2} is then given by
\cite{mp97}
\beq
F={{\gw^4}\over 4} 
{{(\mi^2-m_{\ol{l}\nu}^2)}\over {(m_t\Ga_t)^2}}
\left [{{\mi^2(1-c_{\ol{l}b}^2)+m_{\ol{l}\nu}^2
(1+c_{\ol{l}b})^2}\over 
{(m_{\ol{l}\nu}^2-m_W^2)^2+(m_W\Gamma_W)^2}}\right ]
\label{f1}
\eeq
while the factor $\ol F$ is given by
\beq
\ol{F}={{\gw^4}\over 4}
{{(\mib^2-m_{q\ol{q}'}^2)}\over {(m_t\Gamma_t)^2}}
\left [{{\mib^2(1-c_{q\ol{b}}^2)+m_{q\ol{q}'}^2
(1+c_{q\ol{b}})^2}\over
{(m_{q\ol{q}'}^2-m_W^2)^2+(m_W\Gamma_W)^2}}\right ].
\label{f2}
\eeq
In these expressions,
$m_{\ol{l}\nu }$ is the invariant mass 
of the lepton and neutrino from the $W^+$ decay, 
and $m_{q\ol{q}'}$ is the 2-jet invariant mass 
from the $W^-$ decay.
Also,
$\mi$ is the invariant mass of the lepton, 
neutrino, and $b$ from the $t$ decay, 
while $\mib$ is the 3-jet invariant 
mass from the $\ol t$ decay. 
The width of the $t$ quark is denoted by $\Ga_t$,
while the mass and width of the $W$ boson are 
$m_W$ and $\Gamma_W$.
The quantity $c_{\ol{l}b}$ is the cosine of the angle 
between the lepton and the $b$ in the $W^+$ rest frame,
and $c_{q\ol{b}}$ is the cosine of the angle between 
the light quarks from the $W^-$ and the $\ol{b}$ 
in the $W^-$ rest frame. 
For simplicity,
we have omitted the CKM factors.
To describe the top line shape,
a correction to the denominator $(m_t\Ga_t)^2$ can be introduced 
\cite{d0-peak},
which in principle would involve the modified top dispersion relation.
In terms of four-momenta,
the quantities $F$ and $\ol{F}$ become
\bea
F&=&-4\gw^4 
\fr {(p_\nu \cdot p_b)(p_{\ol{\ell}}\cdot p_t)}
{(m_t\Gamma_t)^2
[(m_{\ol{\ell}\nu}^2-m_W^2)^2 +(m_W\Gamma_W)^2]} ~,
\nonumber\\
\ol{F}&=&-4\gw^4 
\fr {(p_q \cdot p_b)(p_{\ol{q}'}\cdot p_{\ol{t}})}
{(m_t\Gamma_t)^2
[(m_{q\ol{q}'}^2-m_W^2)^2+(m_W\Gamma_W)^2]} .
\nonumber\\
\eea
For other decays of the $t$ and $\ol{t}$,
the corresponding decay-product momenta 
can be substituted appropriately.

\subsection{Lorentz-violating case} 
\label{Lorentz-violating case} 

In the presence of Lorentz violation,
additional Feynman diagrams contribute
to the matrix element
for $t$-$\ol t$ production and decay.
As before,
the narrow-width approximation 
factors the matrix element into a production part 
and two decay parts. 
The additional contributions to the Feynman diagrams 
for each of these parts arise as Lorentz-violating insertions 
on the propagators and vertices,
in accordance with the general discussion given in Sec.\ \ref{Observables}. 
The effects of each type of coefficient
can be considered in turn.

Consider first the prospective contributions 
from the coefficients associated with CPT violation,
which are contained in the Lagrange-density terms \rf{cptviol}
or,
equivalently,
in Eq.\ \rf{cptviolqequiv}.
It turns out that these coefficients produce no relevant effects,
as can be seen in several ways.
A general line of reasoning considers the production process
$q \ol q \to t \ol t$ or $gg \to t \ol t$
at all orders,
assuming no polarizations or spins are detected.
If this process were CPT violating,
then its squared amplitude would necessarily have a contribution 
proportional to an odd power of coefficients for CPT violation
and therefore should change sign under the CPT operation.
However,
the CPT-conjugate squared amplitude can be obtained
by interchanging $q\leftrightarrow \ol q$
for all quarks including the top,
which by inspection of the generic Feynman diagram
yields the same squared amplitude as the original process.
This is impossible unless no CPT violation occurs.

To understand more explicitly why CPT violation
has no relevant effect for unpolarized $t$-$\ol t$ production and decay,
we can consider various specific insertions in turn.
Using the first line of Eq.\ \rf{cptviolqequiv},
for example,
we see that CPT violation involves only right-handed $t$ or $b$ fields.
Inspection reveals that every diagram with a corresponding insertion 
on a $t$ or $b$ propagator
is accompanied by a conjugate diagram yielding a contribution
of the same magnitude but opposite sign.
This cancellation also holds 
in the context of the narrow-width approximation.
We can therefore disregard CPT violation
in $t$-$\ol t$ production and decay
without loss of generality.

Next,
consider effects involving the coefficients associated 
with CPT-even Lorentz-violating operators
given by Eq.\ \rf{lorviol}.
Consider first the coefficient $H_{\mn}$. 
Insertion of the associated operator 
on the production side gives a vanishing matrix element. 
Insertion in the decay diagrams gives a nonzero matrix element,
but the result vanishes when the phase-space integral
is performed for the $t$ and $\ol t$ decays. 
This suggests that insertion of the $H_{\mn}$ operator
generates spin-correlation effects,
which can be neglected for present purposes.
Any real effects of this type could in principle be observed 
as angular information in the final state, 
but they are suppressed by the coefficients for Lorentz violation 
relative to the usual spin correlations
that exist in the Lorentz-invariant 
top-quark production and decay
and hence are unlikely to be candidates for precision measurement.

The Lorentz-violating contributions of interest are therefore
those involving the $c_\mn$-type coefficients. 
To express these effects,
we write the square of the matrix element 
at first order in Lorentz violation in the form
\begin{eqnarray}
|{\cM}|^2 &=&
PF\ol{F}+(\de_p P)F\ol{F}+(\de_v P)F\ol{F}
\nonumber\\
&&
+P(\delta F)\ol{F}+ PF(\delta\ol{F}).
\label{ttbarmatrixsquared}
\end{eqnarray}
The first term is the Lorentz-invariant piece
given in the previous section,
while the other terms represent the corrections
arising at leading order in the $c_\mn$-type coefficients.
The subscripts $p$ and $v$ refer to propagator and vertex insertions,
respectively.
In what follows,
we discuss each of the Lorentz-violating corrections in turn.

\subsubsection{Production via quark fusion} 
\label{Quark fusion} 

For production via quark fusion,
the corrections can conveniently be expressed 
in terms of the coefficient $c_{\mu\nu}$
introduced in Eq.\ \rf{LRdefs}.
The contribution from insertions on the $t$ and $\ol t$ propagators is
\bea
\de_p P&=& 
{\gs^4\over {18E^4}}
c_{\mu\nu}
\Big[
(p_q\cdot p_t)(p_{\ol{t}}^\mu p_{\ol{q}}^\nu)
+(p_q\cdot p_{\ol{t}})
(p_t^\mu p_{\ol{q}}^\nu)
\nonumber \\
&&
+(p_{\ol{q}}\cdot p_t)
(p_{\ol{t}}^\mu p_q^\nu)
+(p_{\ol{q}}\cdot p_{\ol{t}})
(p_t^\mu p_q^\nu)
\Big].
\eea
The contribution from the production vertex is
\bea
\de_v P&=&
{\gs^4\over {18E^4}}
c_{\mu\nu}
\Big [
-(p_q\cdot p_{\ol{q}})(p_t^\mu p_{\ol{t}}^\nu
+p_{\ol{t}}^\mu p_t^\nu) 
\nonumber \\
&&
-(p_t\cdot p_{\ol{t}}+m_t^2)(p_q^\mu p_{\ol{q}}^\nu
+p_{\ol{q}}^\mu p_q^\nu)
\nonumber \\
&&
+(p_q\cdot p_t)p_{\ol{q}}^\mu p_{\ol{t}}^\nu
+(p_q\cdot p_{\ol{t}})p_{\ol{q}}^\mu p_t^\nu
\nonumber \\
&&
+(p_{\ol{q}}\cdot p_t)p_q^\mu p_{\ol{t}}^\nu
+(p_{\ol{q}}\cdot p_{\ol{t}})p_q^\mu p_t^\nu
\Big ].
\eea
The above expressions have been used by the D0 Collaboration
to perform a search for Lorentz violation 
in the top-quark sector
using data from the Fermilab Tevatron
\cite{D0top}.
In principle,
the statistical power of this analysis could be enhanced
by incorporating also the production from gluon fusion
described in the next section,
which would give additional contributions 
suppressed at the 15\% level relative to the ones above
\cite{knk}.

\subsubsection{Production via gluon fusion} 
\label{Gluon fusion} 

For production via gluon fusion,
the correction from the propagators can be separated into six parts,
three coming from squaring the amplitude of each Feynman diagram
and three from the interference between the amplitudes of diagram pairs.
These six contributions represent the dominant ones
for $t$-$\ol t$ production at LHC energies
and hence are well suited for a Lorentz-violation search
using data from the LHC detectors. 

Adopting the same notation as in Eq.\ \rf{mandelstam},
the squared moduli for the $s$, $t$, and $u$ channels
give the three contributions
\begin{widetext}
\bea
\delta _p\ol{\sum}|\cM_{ss}|^2&=&
{{3\gs^4c_{\mu\nu}}\over {4s^2}}\big [s(p_t^\mu p_{\ol t}^\nu+p_{\ol t}^\mu p_t^\nu) 
+ (t-m_t^2)(p_1^\mu p_{\ol t}^\nu+p_2^\mu p_t^\nu)\
+(u-m_t^2)( p_1^\mu p_t^\nu + p_2^\mu p_{\ol t}^\nu)\big ],
\eea
\bea
\delta _p\ol{\sum}|\cM_{tt}|^2&=&
{{\gs^4c_{\mu\nu}}\over {6(t-m_t^2)^3}}
\big [(-t^2+tu-m_t^2 u+3m_t^2 t-10m_t^4)(p_1^\mu p_2^\nu + p_2^\mu p_1^\nu)
\nonumber\\
&&
\hskip 50pt
+(t^2+tu-m_t^2 u-9m_t^2 t)
( p_t^\mu p_{\ol t}^\nu + p_{\ol t}^\mu p_t^\nu )
\nonumber\\
&&
\hskip 50pt
+(-t^2-tu+m_t^2 u+5m_t^2 t+4m_t^4)
(p_1^\mu p_t^\nu +p_t^\mu p_1^\nu+p_2^\mu p_{\ol t}^\nu +p_{\ol t}^\mu p_2^\nu)\big ],
\eea
and 
\bea
\delta _p\ol{\sum}|\cM_{uu}|^2&=&
{{\gs^4c_{\mu\nu}}\over {6(u-m_t^2)^3}}
\big [(-u^2+tu-m_t^2 t+3m_t^2 u-10m_t^4)(p_1^\mu p_2^\nu + p_2^\mu p_1^\nu)
\nonumber\\
&&
\hskip 50pt
+(u^2+tu-m_t^2 t-9m_t^2 u)
( p_t^\mu p_{\ol t}^\nu + p_{\ol t}^\mu p_t^\nu )
\nonumber\\
&&
\hskip 50pt
+(-u^2-tu+m_t^2 t+5m_t^2 u+4m_t^4)
(p_1^\mu p_{\ol t}^\nu +p_{\ol t}^\mu p_1^\nu+p_2^\mu p_t^\nu +p_t^\mu p_2^\nu)\big ].
\eea
The three interference terms yield the contributions 
\bea
\delta _p\ol{\sum}|\cM_{st}|^2&=&
{{3\gs^4c_{\mu\nu}}\over {32s(t-m_t^2)^2}}\big [ \{2 t s -(t+m_t^2)(u-m_t^2)\}
(p_1^\mu p_t^\nu + p_t^\mu p_1^\nu+p_2^\mu p_{\ol t}^\nu + p_{\ol t}^\mu p_2^\nu)
\nonumber \\
&&
\hskip 60pt
+(t-m_t^2)\{(3t-5 m_t^2)
(p_1^\mu p_{\ol t}^\nu + p_{\ol t}^\mu p_1^\nu + p_2^\mu p_t^\nu + p_t^\mu p_2^\nu)
\nonumber \\
&&
\hskip 60pt
+(t+3 u -8 m_t^2)
(p_1^\mu p_2^\nu - p_2^\mu p_1^\nu -p_1^\mu p_{\ol t}^\nu 
+ p_{\ol t}^\mu p_1^\nu +p_2^\mu p_{\ol t}^\nu - p_{\ol t}^\mu p_2^\nu)\}
\nonumber \\
&&
\hskip 60pt
-2\{ 8 m_t^4+(t-3 m_t^2)(3 t+u) \} (p_1^\mu p_2^\nu + p_2^\mu p_1^\nu) 
\nonumber \\
&&
\hskip 60pt
+ 4(2 t u-3 m_t^2 t-m_t^2 u + 2 m_t^4)
( p_t^\mu p_{\ol t}^\nu + p_{\ol t}^\mu p_t^\nu )
\big ],
\eea
\bea
\delta _p\ol{\sum}|\cM_{su}|^2&=&
{{3\gs^4c_{\mu\nu}}\over {32s(u-m_t^2)^2}}\big [ \{2 u s -(u+m_t^2)(t-m_t^2)\}
(p_1^\mu p_{\ol t}^\nu + p_{\ol t}^\mu p_1^\nu+p_2^\mu p_t^\nu + p_t^\mu p_2^\nu)
\nonumber \\
&&
\hskip 60pt
+(u-m_t^2)\{(3u-5 m_t^2)
(p_1^\mu p_t^\nu + p_t^\mu p_1^\nu + p_2^\mu p_{\ol t}^\nu + p_{\ol t}^\mu p_2^\nu)
\nonumber \\
&&
\hskip 60pt
+(u+3 t -8 m_t^2)
(p_1^\mu p_2^\nu - p_2^\mu p_1^\nu -p_1^\mu p_t^\nu 
+ p_t^\mu p_1^\nu +p_2^\mu p_t^\nu - p_t^\mu p_2^\nu)\}
\nonumber \\
&&
\hskip 60pt
-2\{ 8 m_t^4+(u-3 m_t^2)(3 u+t) \}(p_1^\mu p_2^\nu + p_2^\mu p_1^\nu) 
\nonumber \\
&&
\hskip 60pt
+ 4(2 t u-3 m_t^2 u-m_t^2 t + 2 m_t^4)
( p_t^\mu p_{\ol t}^\nu + p_{\ol t}^\mu p_t^\nu )
\big ],
\eea
and 
\bea
\delta _p\ol{\sum}|\cM_{tu}|^2&=&
{{\gs^4c_{\mu\nu}}\over {24(u-m_t^2)^2(t-m_t^2)^2}}
\big [(2s+m_t^2)(t-m_t^2)(u-m_t^2)
(p_1^\mu p_2^\nu + p_2^\mu p_1^\nu
- p_t^\mu p_{\ol t}^\nu - p_{\ol t}^\mu p_t^\nu )
\nonumber\\
&&
\hskip 100pt
+m_t^2\{(s^2-7m_t^2 s-3tu+3m_t^4)
(p_1^\mu p_2^\nu + p_2^\mu p_1^\nu
+ p_t^\mu p_{\ol t}^\nu + p_{\ol t}^\mu p_t^\nu )
\nonumber\\
&&
\hskip 100pt
-(t-m_t^2)(t-u+4m_t^2)
(p_1^\mu p_{\ol t}^\nu +p_{\ol t}^\mu p_1^\nu+p_2^\mu p_t^\nu + p_t^\mu p_2^\nu)
\nonumber\\
&&
\hskip 100pt
+(u-m_t^2)(t-u-4m_t^2)
(p_1^\mu p_t^\nu + p_t^\mu p_1^\nu+p_2^\mu p_{\ol t}^\nu + p_{\ol t}^\mu p_2^\nu)\}\big ],
\eea
arising from the $s$-$t$, $s$-$u$, and $t$-$u$ channel interferences, respectively.

The corrections arising via vertex insertions can similarly be written
as the sum of six terms.
The contributions from the squared moduli of each Feynman diagram are
\bea
\delta _v\ol{\sum}|\cM_{ss}|^2&=&
{{3\gs^4c_{\mu\nu}}\over {4s^2}}
\big [
t (p_{\ol t}^\mu p_1^\nu + p_t^\mu p_2^\nu-p_1^\mu p_2^\nu - p_2^\mu p_1^\nu)
+u(p_{\ol t}^\mu p_2^\nu + p_t^\mu p_1^\nu-p_1^\mu p_2^\nu - p_2^\mu p_1^\nu)
\nonumber\\
&&
\hskip 40pt
-m_t^2((p_1-p_2)^\mu (p_1-p_2)^\nu)\big ],
\eea
\bea
\delta _v\ol{\sum}|\cM_{tt}|^2&=&
{{\gs^4c_{\mu\nu}}\over {3(t-m_t^2)^2}}
\big [
(t-3m_t^2)(p_1^\mu p_t^\nu + p_t^\mu p_1^\nu+p_2^\mu p_{\ol t}^\nu + p_{\ol t}^\mu p_2^\nu)
+4m_t^2
( p_t^\mu p_{\ol t}^\nu + p_{\ol t}^\mu p_t^\nu )
\big ],
\eea
and 
\bea
\delta _v\ol{\sum}|\cM_{uu}|^2&=&
{{\gs^4c_{\mu\nu}}\over {3(u-m_t^2)^2}}
\big [
(u-3m_t^2)(p_1^\mu p_{\ol t}^\nu + p_{\ol t}^\mu p_1^\nu+p_2^\mu p_t^\nu + p_t^\mu p_2^\nu)
+4m_t^2
( p_t^\mu p_{\ol t}^\nu + p_{\ol t}^\mu p_t^\nu )
\big ],
\eea
while the remaining contributions give
\bea
\delta _v\ol{\sum}|\cM_{st}|^2&=&
{{3\gs^4c_{\mu\nu}}\over {32s(t-m_t^2)}}
\big [
2(s+4m_t^2)(p_1^\mu p_2^\nu + p_2^\mu p_1^\nu)
+(4t+3u-13m_t^2)
(p_1^\mu p_t^\nu +p_t^\mu p_1^\nu +p_2^\mu p_{\ol t}^\nu +p_{\ol t}^\mu p_2^\nu)
\nonumber \\
&&
+4(t-u) 
( p_t^\mu p_{\ol t}^\nu + p_{\ol t}^\mu p_t^\nu )
+(-2t-3u+7m_t^2)(p_{\ol t}^\mu p_1^\nu +p_t^\mu p_2^\nu)
+(3u-9m_t^2)(p_1^\mu p_{\ol t}^\nu +p_2^\mu p_t^\nu)\big ],
\eea
\bea
\delta _v\ol{\sum}|\cM_{su}|^2&=&
{{3\gs^4c_{\mu\nu}}\over {32s(u-m_t^2)}}
\big [
2(s+4m_t^2)(p_1^\mu p_2^\nu + p_2^\mu p_1^\nu)
+(4u+3t-13m_t^2)
(p_1^\mu p_{\ol t}^\nu +p_{\ol t}^\mu p_1^\nu +p_2^\mu p_t^\nu +p_t^\mu p_2^\nu)
\nonumber \\
&&
+4(u-t)
( p_t^\mu p_{\ol t}^\nu + p_{\ol t}^\mu p_t^\nu )
+(-2u-3t+7m_t^2)(p_t^\mu p_1^\nu +p_{\ol t}^\mu p_2^\nu)
+(3t-9m_t^2)(p_1^\mu p_t^\nu +p_2^\mu p_{\ol t}^\nu)\big ],
\eea
and 
\bea
\delta _v\ol{\sum}|\cM_{tu}|^2&=&
{{\gs^4c_{\mu\nu}}\over {6(t-m_t^2)(u-m_t^2)}}
\big [
m_t^2\{(p_1^\mu p_2^\nu + p_2^\mu p_1^\nu)
-2 ( p_t^\mu p_{\ol t}^\nu + p_{\ol t}^\mu p_t^\nu ) \}
\big ]
\eea
from the $s$-$t$, $s$-$u$, and $t$-$u$ channel interferences, 
respectively.

Note that the above individual expressions lack manifest symmetry
in the indices $\mu$ and $\nu$,
even though the discussion in Sec.\ \ref{Observables}
reveals that the antisymmetric contribution must be unphysical. 
However,
only the total sum of the contributions from all the Feynman diagrams 
represents a physical observable, 
and the symmetry of this sum can readily be verified.
Note also that 
terms involving the trace $c^\mu_{\pt{\mu}\mu}$ 
can be set to zero
in all the expressions for the production process,
for reasons outlined in Sec.\ \ref{Observables}.

\subsubsection{Semileptonic decay} 
\label{Semileptonic decay} 

For the $t$ and $\ol t$ decays,
the Lorentz-violating effects involve only the coefficient $(c_L)_\mn$,
as can be seen from Eq.\ \rf{lorviol}.
Assuming as before the decay channels \rf{tdecay} and  \rf{tbardecay},
the decay contributions $\de F$ and $\de \ol F$ 
to the matrix element \rf{ttbarmatrixsquared} are given by 
\bea
\delta F&=&
2\gw^4
{1\over {(m_t\Gamma_t)^2}
{[(m_{\ol{l}\nu}^2-m_W^2)^2
+(m_W\Gamma_W)^2]}}
(c_L)_{\mu\nu}
\nonumber \\
&&
\times
\Big [
(p_b\cdot p_t)(p_\nu^\mu p_{\ol{\ell}}^\nu
+p_{\ol{\ell}}^\mu p_\nu^\nu)
+(p_b\cdot p_\nu)(p_t^\mu p_{\ol{\ell}}^\nu
+p_{\ol{\ell}}^\mu p_t^\nu)
-(p_b\cdot p_{\ol{\ell}})(p_t^\mu p_\nu^\nu
+p_\nu^\mu p_t^\nu)
-(p_t\cdot p_\nu)(p_b^\mu p_{\ol{\ell}}^\nu
+p_{\ol{\ell}}^\mu p_b^\nu)
\nonumber \\
&&
\qquad
+(p_t\cdot p_{\ol{\ell}})(p_b^\mu p_\nu^\nu
+p_\nu^\mu p_b^\nu)
+(p_\nu\cdot p_{\ol{\ell}})(p_b^\mu p_t^\nu
+p_t^\mu p_b^\nu)
\Big ]
\eea
and 
\bea
\delta \ol{F}&=&
2\gw^4
{1\over {(m_t\Gamma_t)^2}
{[(m_{\ol{l}\nu}^2-m_W^2)^2
+(m_W\Gamma_W)^2]}}
(c_L)_{\mu\nu}
\nonumber \\
&&
\times
\Big [
(p_{\ol{t}}\cdot p_{\ol{b}})(p_q^\mu p_{\ol{q}'}^\nu
+p_{\ol{q}'}^\mu p_q^\nu)
+(p_{\ol{t}}\cdot p_q)(p_{\ol{b}}^\mu p_{\ol{q}'}^\nu
+p_{\ol{q}'}^\mu p_{\ol{b}}^\nu)
-(p_{\ol{t}}\cdot p_{\ol{q}'})(p_{\ol{b}}^\mu p_q^\nu
+p_q^\mu p_{\ol{b}}^\nu)
-(p_{\ol{b}}\cdot p_q)(p_{\ol{t}}^\mu p_{\ol{q}'}^\nu
+p_{\ol{q}'}^\mu p_{\ol{t}}^\nu)
\nonumber \\
&&
\qquad
+(p_{\ol{b}}\cdot p_{\ol{q}'})(p_{\ol{t}}^\mu p_q^\nu
+p_q^\mu p_{\ol{t}}^\nu)
+(p_q\cdot p_{\ol{q}'})(p_{\ol{t}}^\mu p_{\ol{b}}^\nu
+p_{\ol{b}}^\mu p_{\ol{t}}^\nu)
\Big ].
\eea
\end{widetext}
Note that terms proportional to the trace 
$(c_L)^\mu_{\pt{\mu}\mu}$ 
are disregarded in the above equations
because they can be set to zero without loss of generality,
as described in Sec.\ \ref{Observables}.

\section{Single-top production}
\label{Single-top production}

Given that no leading-order CPT violation appears in $t$-$\ol t$ production,
an interesting issue is whether and how CPT symmetry 
can be studied in the top-quark sector.
In this section,
we address the prospects 
for searches for CPT violation using single-top production.
Although thousands of single top or antitop quarks 
were produced at the Tevatron,
their observation there is challenging
\cite{tevatronsinglet}
and so we focus here on single top or antitop production at the LHC,
where millions of single top or antitop quarks 
are eventually expected to be produced.
Following remarks on the Lorentz-invariant case,
we derive the matrix elements for each of the relevant tree-level channels
and discuss some issues about extracting CPT observables from the LHC dataset.
Our results demonstrate one path to a first test of CPT symmetry 
in the top-quark sector in the context of effective field theory.

\subsection{Lorentz-invariant case} 
\label{Lorentz-invariant case for single top} 

\begin{figure}
\vskip -40pt
\begin{center}
\centerline{\psfig{figure=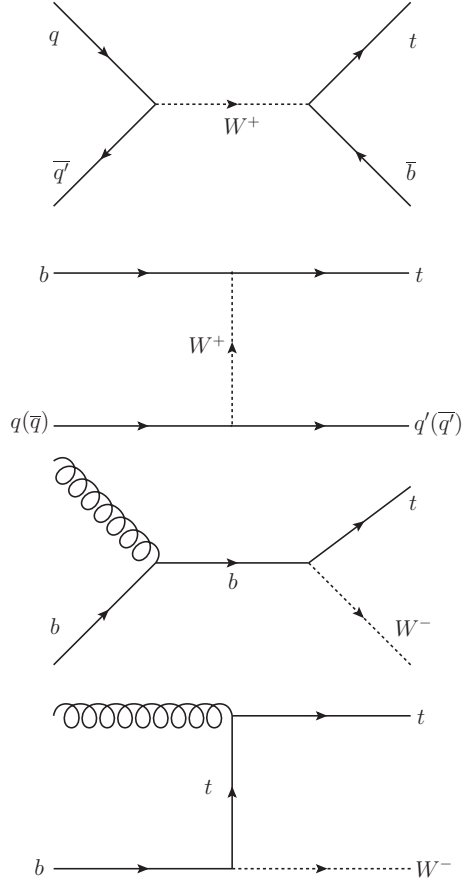,width=1.5\hsize}}
\vskip -100pt
\caption{
\label{fig2}
Tree-level Feynman diagrams for
single-top production via 
the $s$ channel
($q \ol {q'} \to t \ol b$),
the $t$ channels
($b q \to t q'$
and $b \ol q \to t \ol {q'}$),
and associated $t$-$W$ production
($b g \to t W^-$). }
\end{center}
\end{figure}

We are interested in the production and subsequent decay 
of a single top or antitop quark.
Consider first the Lorentz- and CPT-invariant scenario
in the usual SM.
In parallel with the treatment of pair production 
in Sec.\ \ref{Lorentz-invariant case}
using the narrow-width approximation,
the production and decay processes 
for a single top or antitop quark can be viewed 
as contributing distinct factors to the squared modulus
of the matrix element.
The decay factor is independent of the production mechanism,
and examples of its form are given by the expressions 
\rf{f1} and \rf{f2}.
On the production side,
four basic processes contribute 
to the tree-level amplitude for single-top production:
$q\ol q$ annihilation via $W$ in the $s$ channel
\cite{schan},
$b q$  and $b \ol q $ weak interactions 
in the $t$ channel
\cite{tchan},
and $b$-gluon production of $tW$
\cite{wchan}.
In each case,
the production factor is the squared matrix element
for the process.
The tree-level diagrams contributing to these processes
are shown in Fig.\ \ref{fig2}.

Averaging over color and spin in the initial state
and summing over color and spin in the final state,
the squared matrix element
for single-top production in $q \overline{q}$ annihilation 
via $W^+$ in the $s$ channel
is given by
\beq
\ol{\sum} |\cM_{q\ol q}|^2 = 
\frac 1 4 {\gw^4} 
|V_{tb}|^2|V_{q q'}|^2
\fr{u(u-m_t^2)} {(s-m_W^2)^2} .
\label{schanm}
\eeq
In this expression,
$V_{tb}$ and $V_{q q'}$ 
are elements of the CKM matrix,
$m_t$ is the mass of the top quark,
$m_W$ is the mass of the $W^+$ boson, 
and the Mandelstam variables $s$, $t$, $u$
are given by 
\bea
s=(p_1+p_2)^2=(p_3+p_4)^2 ,
\nonumber \\
t= (p_2-p_4)^2 = (p_1-p_3)^2 ,
\nonumber \\
u=(p_1-p_4)^2=(p_2-p_3)^2 ,
\label{mv}
\eea
with $1,2,3,4=\overline{q'},q,\overline{b},t$.
Contributions to Eq.\ \rf{schanm}
from CKM-suppressed processes
can also occur but are disregarded here for simplicity.

For single-top production via $bq$ weak interaction in the $t$ channel,
the Lorentz-invariant squared matrix element is found to be 
\beq
\ol{\sum} |\cM_{bq}|^2 =
\frac 1 4 {\gw^4} 
|V_{tb}|^2|V_{q q'}|^2
\fr {s(s-m_t^2)} {(t-m_W^2)^2},
\eeq
where the Mandelstam variables \rf{mv}
are defined with $1,2,3,4=q,b,q',t$.
The analogous result
for single-top production via $b\ol{q}$ weak interaction 
in the $t$ channel is
\beq
\ol{\sum} |\cM_{b\ol q}|^2 =
\frac 1 4 {\gw^4} 
|V_{tb}|^2|V_{q q'}|^2
\fr {u(u-m_t^2)} {(t-m_W^2)^2} ,
\eeq
where now $1,2,3,4=\ol{q},b,\ol{q'},t$.

The fourth process,
$bg$ associated production of $tW$,
acquires contributions from the last two diagrams in Fig.\ \ref{fig2}.
Including interference terms between the two diagrams,
the squared matrix element is given in the SM by
\bea
\ol{\sum} |{\cM_{bg}}|^2 &=&
\dfrac{\gw^2 \gs^2}{24} |V_{tb}|^2
\bigg\{ -\dfrac{2 m_t^2}{m_W^2}
- \bigg( \dfrac{m_t^2}{m_W^2}+2  \bigg)
\nonumber\\
&&
\hskip -35pt
\times
\bigg[ 
\dfrac{s}{t-m_t^2}+ \dfrac{m_t^2-2 m_W^2+t}{s}
\nonumber\\
&&
\hskip -25pt
+\dfrac{2(m_t^2-m_W^2)}{t-m_t^2} 
\bigg( \dfrac{m_t^2}{t-m_t^2}+\dfrac{m_t^2-m_W^2}{s}+1 \bigg) 
\bigg] 
\bigg\},
\nonumber\\
\eea
where the Mandelstam variables \rf{mv}
are defined using $1,2,3,4=b,g,W,t$.

The situation for single-antitop production
can be studied in a similar way.
The relevant diagrams for the available tree-level processes
can be obtained by changing the charges on the $W$ bosons
and reversing the direction
of all the fermion lines in Fig.\ \ref{fig2}.
The resulting squared matrix elements have the same forms
as those given above,
reflecting the CPT invariance of the SM. 

At LHC energies,
the cross section for single-top production in the $s$ channel 
is several times smaller than that for the $tW$ mode
and more than an order of magnitude smaller 
than the dominant $t$ channels
\cite{wb}.
Note that at the LHC
the SM cross sections for the $tW$ and the $\ol t W$ modes 
are equal by CPT invariance.
However, 
the cross section for single-$t$ production
in either the $s$ or the $t$ channels
is larger than that for single-$\ol t$ production 
because the LHC is a proton-proton collider.
The CPT transformation in these channels relates results for the LHC
to those for a hypothetical `anti-LHC' 
involving antiprotons colliding with antiprotons. 

The $s$-channel has features similar to 
the quark-annihilation $s$-channel calculation 
for $t$-$\ol t$ production. 
However,
calculation of the cross sections for the $t$ channels and $tW$ mode
faces a technical obstacle.
The $b$ quark involved in these channels arises 
as part of the quark-gluon sea of the colliding proton,
so even in the SM the diagrams shown in Fig.\ \ref{fig2}
are insufficient to yield accurate predictions
for the cross sections. 
The $b$ quark can be viewed as emerging
from $g \to b \ol b$ pair production in the sea,
so in effect the $t$ channel involves 
$g q \to t q' \ol b$
while the $tW$ mode involves
$g g \to t W^- \ol b$.
In these processes,
when the $b$ quark moves collinear to its parent gluon, 
a divergence appears 
that is regulated by the $b$-quark mass. 
The effective perturbation expansion then contains large logarithms
involving 
$\al_{\rm QCD} \ln(m_W/m_b)$ and $\al_{\rm QCD} \ln(m_t/m_b)$ 
instead of the usual $\al_{\rm QCD}$,
so standard perturbation theory is unreliable. 
Instead,
the $b$ quark must be handled via a parton distribution function
obtained from perturbative QCD using gluons 
and parton distribution functions for the light quarks
\cite{acot}.
Evolving the parton distribution function for the $b$-quark 
using the Altarelli-Parisi equation sums the large logarithms
and enables calculation of the cross sections for these channels.

\subsection{CPT-violating case} 
\label{CPT-violating case} 

Since single-top and $t$-$\ol t$ production 
involve distinct processes,
it is reasonable to expect that studying the former
would provide access to additional SME coefficients.
In particular,
as the results of Sec.\ \ref{Top-antitop pair production}
show that CPT violation is inaccessible in $t$-$\ol t$ production,
single-top processes are of substantial potential interest
for studies of CPT symmetry in the top-quark sector.

To investigate this possibility in a direct and comparatively simple way,
we can restrict the SME Lagrange density for the top quark
presented in Sec.\ \ref{SME Lagrange density for the top}
to the special case 
where all coefficients for Lorentz violation are set to zero 
except those controlling CPT violation in the top-quark sector.
Adopting the field redefinitions leading to Eq.\ \rf{cptviolqequiv},
this corresponds to allowing only CPT violation
involving the right-handed top quark,
with the sole nonzero coefficient 
then being the coefficient $b_\mu$ introduced in Eq.\ \rf{LRdefs}.
In this simplified model,
all leading-order CPT-even Lorentz violation is absent.
Also,
CPT-violating effects are limited 
to right-handed contributions to the $t$ propagator.

Paralleling the CPT-invariant case
described in Sec.\ \ref{Lorentz-invariant case for single top}, 
the squared modulus $|\cM|^2$ of the amplitude can be written 
in the narrow-width approximation
as the product of a production factor and a decay factor.
The decay factor is comparatively straightforward to handle 
within the above scenario,
since the only possible effect on the decay Feynman diagrams 
arises from a right-handed insertion on the initial $t$ propagator.
However,
direct calculation reveals that the CPT-violating contribution
to the squared matrix element vanishes,
in analogy with the zero contribution from $H_\mn$
discussed in Sec.\ \ref{Lorentz-violating case}.
The decay factor for single-top production
is therefore independent of CPT violation
and can be taken to have a conventional form.

Determining the contributions to the production factor
requires more effort.
Within the above assumptions,
the CPT-violating processes 
in the tree-level Feynman diagrams for single-top production 
involve right-handed insertions on the $t$ propagators
in the diagrams shown in Fig.\ \ref{fig2}.
At leading order,
only one insertion is allowed at a time,
yielding a total of five diagrams to consider. 
The corrections to the corresponding matrix element at leading order
therefore involve the interference terms
between these five diagrams and the Lorentz-invariant ones 
in Fig.\ \ref{fig2}.
For example,
the two Lorentz-invariant diagrams for the $tW$ mode
are supplemented with three CPT-violating diagrams,
each having an insertion on a $t$-quark line,
so the corrections to the matrix element for this process 
contain six terms at leading order. 

Several calculational simplifications emerge 
in the evaluation of the various contributions 
to the production matrix elements.
Note first that in all diagrams for production the $t$ propagators 
are connected to a left-handed weak flavor-changing vertex.
This means the right-handed $t$-quark insertions
are partially cancelled by the projections,
which reduces the complexity of some calculations.
Another factor of relevance for simplifications 
is the ratio $m_b/m_t\simeq 1/40$,
which means it is reasonable to neglect the $b$-quark mass 
in the calculations.

One complication appearing in the calculation 
of the squared matrix element is the determination of the spin sums
over the top-quark states.
The presence of CPT violation modifies these sums 
compared to the results for a conventional Dirac spinor.
For the scenario of interest here,
we find 
\bea
\sum\limits_{\al=1,2} {u^{(\al)} \ol{u}^{(\al)}} &=&
\slashed{p}+m_t+\slashed{b}-\fr{p \cdot b}{m_t^2}(1+\ga_5)\slashed{p},
\nonumber\\
\sum\limits_{\al=1,2} {v^{(\al)} \ol{v}^{(\al)}} &=&
\slashed{p}-m_t-\slashed{b}+\fr{p \cdot b}  {m_t^2}(1+\ga_5)\slashed{p},
\label{spinsum}
\eea
where $u^{(\al)}(p)$ and $v^{(\al)}(p)$ are the eigenspinors
of the modified Dirac equation.
The derivation of this result is outlined 
in Appendix \ref{appendix}. 

With the above considerations,
the calculation of the leading-order CPT-violating corrections 
to the squared matrix elements for the various single-top production processes
can proceed in a straightforward manner.
After some calculation, 
we find that the correction to the squared matrix element
for single-top production in $q \overline{q}$ annihilation 
via $W^+$ in the $s$ channel
takes the form
\beq
\de \ol{\sum} |\cM_{q\ol q}|^2 = 
-\frac 1 4 {\gw^4} 
|V_{tb}|^2|V_{q q'}|^2
\fr{2b \cdot p_1} {(s-m_W^2)^2} ,
\label{t-s}
\eeq
where the Mandelstam variables \rf{mv}
are defined with $1,2,3,4=\ol{q'},q,\ol{b},t$
and contributions from CKM-suppressed analogues are disregarded
as before.
To obtain this result,
we have averaged over color and spin in the initial state
and summed over color and spin in the final state,
as in the CPT-invariant case.

Similar calculations for single-top production via $bq$ 
and via $b\ol{q}$ weak interactions in the $t$ channel
reveal that the corresponding CPT-violating corrections
to the squared matrix elements are
\beq
\de \ol{\sum} |\cM_{bq}|^2 =
\frac 1 4 {\gw^4} 
|V_{tb}|^2|V_{q q'}|^2
\fr {2b \cdot p_3} {(t-m_W^2)^2},
\label{bq}
\eeq
where $1,2,3,4=q,b,q',t$,
and
\beq
\de \ol{\sum} |\cM_{b\ol q}|^2 =
-\frac 1 4 {\gw^4} 
|V_{tb}|^2|V_{q q'}|^2
\fr {2b \cdot p_1} {(t-m_W^2)^2} ,
\label{bqb}
\eeq
where $1,2,3,4=\ol{q},b,\ol{q'},t$.

Finally,
we can obtain the leading-order CPT-violating correction
to $bg$ associated production of $tW$,
which as mentioned above
arises from the interference of three CPT-violating amplitudes
with two Lorentz-invariant ones.
The correction to the squared matrix element in this case is given by
\begin{widetext}
\bea
\de \ol{\sum} |{\cM_{bg}}|^2 &=&
-\fr{\gw^2 \gs^2}{12} 
|V_{tb}|^2
b \cdot \bigg\{ 
\fr {1}{m_W^2 s}
\big[p_2(m_t^2-2m_W^2)+p_3 m_t^2+ p_4 t\big]
+\fr{8m_t^4}{(t-m_t^2)^3} 
\bigg( \fr{m_t^2}{m_W^2}p_3-p_3-p_1\bigg)
\nonumber \\ 
&&
+\fr{1}{m_W^2 s (t-m_t^2)} 
\big[ p_1(2m_t^4-4m_W^4-m_t^2 s)+p_2(3 m_t^4 - 5 m_t^2 m_W^2)
\nonumber \\ 
&&
\hskip 100pt
+(s+m_t^2-m_W^2)(4m_t^2 p_3 -s p_4) + m_W^2 s (2p_3+p_4) 
\big] 
\nonumber \\
&&
+\fr{m_t^2}{(t-m_t^2)^2}
\bigg[ p_1\bigg( \fr{3m_t^2}{m_W^2}-3 \bigg) 
+ 2 p_2 \bigg( \fr{m_t^4}{m_W^2 s}-\fr{m_t^2}{s} 
+ \fr{s}{m_W^2}-\fr{m_W^2}{m_t^2} \bigg) 
\nonumber \\
&&
+ p_3 \bigg( \fr{4m_t^4}{m_W^2 s}+\fr{8m_W^2-12m_t^2}{s}
+\fr{9m_t^2}{m_W^2}-13 \bigg) 
+ p_4 \bigg( \fr{4m_W^2-4m_t^2}{s}-\fr{m_t^2+3s}{m_W^2}
+\fr{2s}{m_t^2}-5 \bigg) 
\bigg]
\bigg\} ,
\label{bg}
\eea
\end{widetext}
where now $1,2,3,4=b,g,W,t$.

The results for single-antitop production
can be obtained in an analogous manner.
The Feynman diagrams for the corresponding processes
can be obtained by changing the charges on the $W$ bosons,
reversing the direction of all the fermion lines 
in Fig.\ \ref{fig2},
and inserting the CPT-violating factor on the various $\ol t$ propagators
in turn.
Since the insertion involves a factor of $-b_\mu$,
and since the leading-order contributions to
the squared matrix elements arise through interference
with the CPT-invariant SM diagrams,
the corrections for single-antitop production
are the negatives of those for single-top production
given in Eqs.\ \rf{t-s}-\rf{bg}.

According to the discussion in Sec.\ \ref{Signals},
the CPT-violating cross section 
for each process leading to single-top or single-antitop production 
exhibits sidereal variations
that can in principle be used to extract constraints
on components of the coefficient $b_\mu$ for CPT violation.
At the LHC,
the cross sections for single-top production 
in any one of the $s$ or $t$ channels 
is different from that for single-antitop production in the same channel
because the LHC is a proton-proton collider
and the light-quark content of the system changes under CPT.
However,
since the $tW^-$ and $\ol t W^+$ modes 
arise from $b$ and $\ol b$ quarks 
that in the SM are equally represented in the sea,
the cross sections for the $tW^-$ and $\ol t W^+$ modes are equal
in the SM. 
As a consequence,
a comparison of the two cross sections
via an asymmetry $\cA_{\rm CPT}$ of the form \rf{asym}
provides a distinct type of sensitivity to CPT violation.
This asymmetry has both a time-independent piece
and a sidereally varying piece,
so careful study of its properties
can provide information about the components $b_Z$ and $b_T$
that cannot readily be accessed 
via sidereal studies of any single process alone.

We remark in passing that 
all the production processes involve the weak interactions,
and hence the single top or antitop quarks can be expected
to emerge with a strong degree of polarization.
An experimental analysis taking this into account
might in principle achieve an enhanced sensitivity to CPT violation.
However,
we have shown above that interesting signals are already present 
in the comparatively simple spin-summed production rates.
These therefore suffice to obtain a first measurement 
of the coefficient $b_\mu$ for CPT violation
using the expected LHC statistics. 

In a more general analysis using all the CPT-violating terms
in Eq.\ \rf{cptviolqequiv},
additional contributions from right-handed insertions
on the $b$-quark propagators could also be included
in the calculations.
Most of the extra corrections arise in a straightforward way,
generating numerous additional interference terms
in the squared modulus of the matrix element.
One additional complication arises in this more general case
because any incoming $b$ quark in the production diagrams
arises from the gluon sea via pair production. 
It is plausible {\it a priori} 
that including Lorentz and CPT violation in the $b$ sector 
affects the parton distribution function for the $b$ quark,
which could require determining the CPT-violating corrections
to the Altarelli-Parisi equation. 
Note, however, that at leading order 
the CPT-violating contributions to the $g\to b\ol b$ vertex 
appear in equal and opposite pairs
because for every CPT-odd insertion associated with the $b$ line
there is an equal and opposite contribution 
associated with the $\ol b$ line.
This effect leads to cancellations in contributions 
from the quark-gluon sea of neutral mesons
in the context of CPT violation in meson oscillations 
\cite{aksidereal}.
It suggests, 
for example,
that in a narrow-width approximation 
for the $b$ production from the sea
it may suffice for some data analyses
to use the Lorentz-invariant parton distribution function
for the $b$ quark in evaluating the CPT-violating contribution
to the amplitude for the $tW$ channel.
A detailed investigation of this intriguing issue 
lies outside our present scope
but would be of definite interest for future work.

\section{Summary and Discussion}
\label{Summary}

In this work,
we investigated the prospects for tests of Lorentz and CPT invariance
in the top-quark sector.
The basic theory is introduced in Sec.\ \ref{Theory}.
Relevant terms involving the top-quark field
are extracted from the general SME framework
describing Lorentz and CPT violation using effective field theory
and are presented in Eqs.\ \rf{lorviol} and \rf{cptviol}. 
The issue of field and coordinate redefinitions
is addressed in Sec.\ \ref{Observables},
and a convenient choice limiting CPT violation
to right-handed fields is presented in Eq.\ \rf{cptviolqequiv}.
Prospective signals to be sought
are discussed in Sec.\ \ref{Signals},
including both sidereal variations for Lorentz and CPT violation 
and asymmetries for CPT-violating rates.

The main results relevant to Lorentz violation 
in $t$-$\ol t$ pair production and decay are discussed 
in Sec.\ \ref{Top-antitop pair production}.
Both pair production by quark fusion and by gluon fusion 
are considered.
The squared modulus of the matrix element for each case
in the Lorentz-invariant limit
is presented explicitly in the narrow-width approximation,
where it factors according to Eq.\ \rf{m2}.
In Sec.\ \ref{Lorentz-violating case},
CPT symmetry is shown to be preserved 
in $t$-$\ol t$ production and decay 
under the theoretical assumptions adopted in this work.
The Lorentz-violating contributions
to the squared modulus of the matrix element
takes the form \rf{ttbarmatrixsquared}.
The Lorentz-violating corrections due to production via quark fusion
are presented in Sec.\ \ref{Quark fusion},
while those for gluon fusion can be found in Sec.\ \ref{Gluon fusion}.
The contributions on the decay side
are obtained in Sec.\ \ref{Semileptonic decay}.

The issue of how to study CPT symmetry in the top-quark sector  
is addressed in Sec.\ \ref{Single-top production}.
We show that single-top and antitop production offers
interesting prospects to search for CPT violation,
and we identify a limiting model for which
calculations of the matrix element are simplified. 
Comments on the Lorentz-invariant case are provided 
in Sec.\ \ref{Lorentz-invariant case},
while the contributions to CPT violation
for the various processes for single-top and single-antitop production
are derived in Sec.\ \ref{CPT-violating case}.

Overall,
the prospects appear good for studying Lorentz and CPT symmetry 
with the top quark.
Already the D0 Collaboration has achieved a sensitivity of about 10\% 
to components of the dimensionless SME coefficients
$(c_L)_{\mu\nu}$ and $(c_R)_{\mu\nu}$
for CPT-even Lorentz violation,
using a sidereal analysis of data from the Fermilab Tevatron.
Our derivation of the matrix element
for $t$-$\ol t$ production via gluon fusion 
given in Sec.\ \ref{Gluon fusion}
now opens the door to a similar analysis at the LHC.
Since the number of $t$-$\ol t$ pairs produced at the LHC
is about an order of magnitude greater than that at the Tevatron,
the attainable sensitivity to 
components of $(c_L)_{\mu\nu}$ and $(c_R)_{\mu\nu}$
can be expected to be of the order of a few percent.

Our proposed methodology for studying CPT symmetry
via single-top production,
described in Sec.\ \ref{CPT-violating case},
suggests access to the CPT-odd coefficient $b_\mu$ for the top quark
via the LHC dataset now lies within reach for the first time.
Since the statistical power in $t$-$\ol t$ production
is around double that for single-top or antitop production,
it is plausible that 
the observer-invariant dimensionless ratio 
$b_\mu p^\mu/s$
could be measured to around 5\%.
The relevant energy scale is set 
by the energy of the initial particles in the production process.
These considerations suggest that the coefficient $b_\mu$,
which has dimensions of mass,
could be measured at a precision of order 100 GeV.
A sidereal study would provide access to the components
$b_X$ and $b_Y$ in the Sun-centered frame,
while a comparison of single-top events with single-antitop events
would give access to $b_T$ and $b_Z$.

The above crude estimate reveals that CPT violation in the top sector
on a scale comparable to the $t$-quark mass
remains a realistic experimental possibility.
Comparatively large but experimentally viable Lorentz violation,
known as countershaded Lorentz violation
\cite{kt},
is typically linked to weaker interactions
that make it challenging to detect.
Countershading is of theoretical interest 
in the context of the Lorentz hierarchy problem
\cite{ksp}
because it has the potential to associate
the scale of Lorentz violation to SM scales 
instead of ones suppressed by a power of the ratio
of the weak scale to the Planck scale.
Countershaded models include,
for example,
ones with Lorentz violation that appears only in matter-gravity couplings
\cite{kt}
or in the pure-gravity sector
\cite{bkx}
and hence is hidden by the comparatively feeble nature 
of the gravitational interaction,
and ones in the neutrino sector
involving oscillation-free Lorentz and CPT violation 
\cite{dkl}
that is hidden by the weak neutrino interactions.

For the top quark,
the existence of countershaded CPT violation remains experimentally plausible 
due to the intrinsic difficulty of measuring 
properties of the top quark via single-top and single-antitop production,
which in turn is a consequence of the weak interactions involved.
Whether this scenario is realized in nature remains to be determined,
but in any case it is evident that the top quark offers
an intriguing open arena for future exploration 
of foundational properties of quantum field theory and the SM,
including in particular Lorentz and CPT symmetry.

\section*{Acknowledgments}

This work is supported in part
by Department of Energy under grant number {DE}-SC0010120
and by the Indiana University Center for Spacetime Symmetries
(IUCSS).

\appendix

\section{Modified spin sums}
\label{appendix}

In this appendix,
we outline the derivation of the modified spin sums
used in Sec.\ \ref{CPT-violating case}
in the derivation of the squared matrix elements
for single-top production in the presence of CPT violation.
The relevant terms in the Lagrange density 
are given by Eq.\ \rf{cptviolqequiv}
with $(a_L)_\mu$ set to zero,
so all the CPT violation
is controlled by the coefficient $b_\mu$.
Leading-order solutions to the resulting modified Dirac equation
can be obtained from the equations in Appendix A 
of the first paper in Ref.\ \cite{ck}
via the substitution $a_\mu \to -b_\mu$.

The spin sum can be readily calculated 
in the zero-momentum frame $S'$.
In what follows,
variables with primes denote quantities in this frame.
The spinors $u^{(\al)}(\vec{p}=0)$ and $v^{(\al)}(\vec{p}=0)$ 
have eigenenergies given to second order in Lorentz violation by
\bea
E^{(\al)\prime}_u &=&
m_t+(-1)^{\al}|\vec{b'}|-b'_0 
+\fr{1}{2m_t}\big[b'_0-(-1)^\al|\vec{b'}|\big]^2 ,
\nonumber\\
E^{(\al)\prime}_v &=&
E^{(\al)\prime}_u 
\big\vert_{b'_0\to -b'_0}.
\nonumber\\
\label{eigenen}
\eea
The explicit forms of $u^{(\al)}(\vec{p}=0)$ and $v^{(\al)}(\vec{p}=0)$ 
at leading order in Lorentz violation are 
\bea
u^{(\al)}(\vec{p}=0) & = & N_u^{(\al)} 
\left( \begin{array}{c}
\ph_u^{(\al)} \\
X_u^{(\al)} \ph_u^{(\al)}
\end{array} \right) ,
\nonumber \\
v^{(\al)}(\vec{p}=0) & = & N_v^{(\al)} 
\left( \begin{array}{c}
X_v^{(\al)} \ph_v^{(\al)}
\\
\ph_v^{(\al)}
\end{array} \right) .
\label{general}
\eea
In these expressions,
the two-component spinors $\ph_{u,v}^{(\al)}$ take the form 
\beq
\ph_{u,v}^{(\al)}=
(\vec{\ka}_{u,v}^{(\al)}\cdot\vec{\si}+\et_{u,v}^{(\al)})
\left(
\begin{array}
{c}0 \\ 1
\end{array}
\right),
\eeq
with 
\bea
{\vec{\ka}}_{u}^{(\al)} &=&
-4m_t \vec{b'}\big[m_t-b'_0+(-1)^{\al}|\vec{b'}|\big],
\nonumber\\
{\vec{\ka}}_{v}^{(\al)} &=&
-{\vec{\ka}}_{u}^{(\al)} 
\big\vert_{b'_0\to -b'_0},
\nonumber\\
\et_{u,v}^{(\al)} &=& -(-1)^{\al}|{\vec{\ka}}_{u,v}^{(\al)}|,
\eea
where we have assumed that
the components of $b'_\mu$ are all smaller than $m_t$.
Also,
the matrices $X_{u,v}^{(\al)}$ in Eq.\ \rf{general} 
are given by 
\beq
X_{u}^{(\al)} = - X_{v}^{(\al)} =  
\fr{1}{2m_t} (\vec{b'}\cdot \vec{\si}-b'_0).
\eeq
For the solutions \rf{general},
we adopt the normalization conditions  
\beq
\ol{u}^{(\al)}u^{(\al)}=- \ol{v}^{(\al)}v^{(\al)}=2m_t,
\eeq
which imply
\beq
|N_{u,v}^{(\al)}|^2=\dfrac{2m_t}{\ph_{u,v}^{(\al)\dagger}\ph_{u,v}^{(\al)}}
\eeq
at leading order.

Using the above results 
to evaluate the spin sums in the frame $S'$ yields 
\bea
\sum\limits_{\al=1,2}u^{(\al)}\ol{u}^{(\al)} &=&
\left(
\begin{array}{cc}
2 m_t & b'_0-\vec{b'}\cdot \vec{\si} \\
-b'_0+\vec{b'}\cdot \vec{\si} & 0
\end{array}
\right),
\nonumber\\
\sum\limits_{\al=1,2}v^{(\al)}\ol{v}^{(\al)} &=&
\left(
\begin{array}{cc}
0 & -b'_0+\vec{b'}\cdot \vec{\si} \\
b'_0-\vec{b'}\cdot \vec{\si} & -2 m_t
\end{array}
\right).
\qquad
\label{spinsums}
\eea
Unlike the Lorentz-invariant case,
the difference of these two results 
is no longer proportional to the identity matrix,
although each of the two spin sums 
still acts as a projection operator on its own subspace.
However,
the completeness relation between the two subspaces
is guaranteed to hold only for the original hamiltonian,
prior to the reinterpretation of negative-energy states.
The CPT-violating shifts reflected in the eigenenergies \rf{eigenen}
introduce nonorthogonal behavior of the two subspaces
upon reinterpretation
\cite{ck}.

To obtain results valid for the frame $S$ in which the spinors
have nonzero momentum $\vec p$,
we can perform an observer Lorentz transformation.
For rapidity $\vec\ze$ in $S$,
the boost takes the form
\beq
u^{(\al)}=Su'^{(\al)},
\quad
v^{(\al)}=Sv'^{(\al)},
\quad
S=\exp(i \ze^{j} \si^{0j}/2),
\quad
\eeq
where as usual
$\si^{0j}=i[\ga^0,\ga^j]/2$.
Under the observer transformation,
the coefficients $b'_\mu$ are related to those in the frame $S$ by
\beq
b'_0=
\ga(b_0-\vec{v}\cdot \vec{b}),
\quad
\vec{b'}=
\vec{b}+\dfrac{\ga-1}{v^2}(\vec{v} \cdot \vec{b})\vec{v}-\ga b_0\vec{v},
\quad
\eeq
where $\vec{v}$ is the velocity of the particle in $S$ 
and $\ga=1/{\sqrt{1- \vec v^2}}$, 
as usual. 
Note that the relationship between $\vec p$ and $\vec v$ 
acquires corrections involving the coefficients for Lorentz violation
\cite{ck},
but at leading order this has no effect on the present derivation. 
Note also that the transformations 
of the spinors and of the coefficients commute.
Implementing these calculations
leads to the modified spin sums \rf{spinsum},
which hold in the frame $S$.

\end{document}